\documentclass[twocolumn,usenatbib]{mnras}

\usepackage{graphicx}
\usepackage{amsmath}

\def\pmb#1{\setbox0=\hbox{#1}%
    \kern-.025em\copy0\kern-\wd0
    \kern.05em\copy0\kern-\wd0
    \kern-.025em\raise.0433em\box0}

\def\ltsima{$\; \buildrel < \over \sim \;$}
\def\gtsima{$\; \buildrel > \over \sim \;$}
\def\simlt{\lower.5ex\hbox{\ltsima}}
\def\simgt{\lower.5ex\hbox{\gtsima}}
\def\p2Y{\;_2Y}
\def\m2Y{\;_{-2}Y}

\def\mk2{\mu {\rm K}^2}
\def\Planck{\it Planck\rm}

\def\LCDM{$\Lambda$CDM}
\def\Dperp{${\cal D}^{\rm perp}$}
\def\Dpar{${\cal D}^{\rm par}$}

\newcommand{\Mpc}{\text{Mpc}} 
\newcommand{\Hunit}{~\text{km}~\text{s}^{-1} \Mpc^{-1}}
\newcommand{\camspec}{{\tt CamSpec}}

\def\pmb#1{\setbox0=\hbox{#1}%
     \kern-.025em\copy0\kern-\wd0
     \kern.05em\copy0\kern-\wd0
     \kern-.025em\raise.0433em\box0}

\begin{document}

\title[BAO and \LCDM ]{ Baryon Acoustic Oscillations from a Different Angle}

\author[George Efstathiou]{George Efstathiou\\
 Kavli Institute for Cosmology Cambridge and 
Institute of Astronomy, Madingley Road, Cambridge, CB3 OHA.}

\maketitle

\begin{abstract}
This paper presents an alternative way of analysing Baryon Acoustic Oscillation (BAO) distance measurements via rotations to define new quantities \Dperp\ and \Dpar. These quantities allow simple tests of consistency with the \Planck\ \LCDM\ cosmology. The parameter  \Dperp\ is determined with negligible uncertainty from \Planck\ under the assumption of \LCDM. Comparing with measurements from the Dark Energy Spectroscopic Instrument (DESI), we find 
that the measurements of \Dperp\ from Data Release 2 (DR2) move into significantly better agreement with the \Planck\ \LCDM\ cosmology compared to
DESI Data Release 1 (DR1). The quantity in the orthogonal direction 
\Dpar\ provides a measure of the physical matter density $\omega_m$ in the \LCDM\ cosmology.  The  DR2 measurements of \Dpar\ remain consistent with \Planck\ \LCDM\ despite the substantial improvement in their accuracy compared to the earlier DR1 results. From the comparison of \Planck\ and DESI BAO measurements, we find no significant evidence in support of evolving dark energy. We also investigate a rotation in
the theory space of the $w_0$ and $w_a$ parameterization of the dark energy equation-of-state $w(z)$. We show that the combination of DESI BAO measurements and the CMB constrain  $w(z=0.5) = -0.996 \pm 0.046$, i.e. very close to the value expected for a cosmological constant. We present a critique of the statistical methodology employed by the DESI collaboration and argue that it gives a misleading impression of the evidence in favour of evolving dark energy.
An Appendix shows that the cosmological parameters determined from the Dark Energy Survey 5 Year supernova sample are in tension with those from
 DESI DR2  and parameters determined by \Planck. 
    
\end{abstract}

\begin{keywords}
cosmology: cosmological parameters, dark energy, supernovae
\end{keywords}

\section{Introduction}
\label{sec:Introduction}

Baryon acoustic oscillation (BAO) features in the galaxy distribution  have provided a powerful geometrical test of the 
background expansion history of the Universe since they were first detected  (\cite{Cole:2005, Eisenstein:2005}, for a review see \cite{Weinberg:2013}). BAO measurements have played a critical role in the development of the \LCDM\ cosmology in the intervening 20 years and   have  become more accurate as galaxy surveys have increased in size. In addition,  the redshift reach has increased to beyond $z=2$ through measurements of BAO features in the distribution of quasars and quasar Ly$\alpha$ absorption lines. BAO 
are complementary to observations of anisotropies in the
cosmic microwave background radiation (CMB). In particular, 
BAO  measurements break geometrical parameter degeneracies in extensions of the \LCDM\ cosmology which  include  spatial curvature and evolving dark energy. The analysis presented in \cite{Planck_Params_2018}
combined \Planck\ CMB data with BAO  results from the completed Sloan Digital Sky Survey Baryon Oscillation Spectroscopy Survey (BOSS)
\citep{Alam:2017} finding consistency with the 
base 6-parameter \LCDM\ cosmology, with no evidence for
spatial curvature or evolution in the equation-of-state of
dark energy. 

However, BAO measurements  from the first year
of observations with the Dark Energy Spectroscopic Instrument (DESI)  have, it is claimed,  provided intriguing  indications of evolving  dark energy (EvDE\footnote{We use the acronym EvDE for evolving dark energy, to avoid confusion with the acronym EDE for  early dark energy. For a review of EDE see e.g. \cite{Poulin:2023}.} )\citep{DESI:2024}. These hints are still present in the three year DESI BAO results \citep{DESI:2025}. Henceforth we will refer to the these two DESI papers as DESI-DR1 and DESI-DR2.  The question of whether the evidence for EvDE from BAO has {\it strengthened} between DR1 and DR2 is addressed in this paper.

The existence of dark energy poses a fundamental problem in physics and remains unexplained \citep[see e.g. the review by][]{Weinberg:1989}. A discovery of EvDE would  have profound implications for physics and could perhaps provide a path towards a theoretical understanding the origin of dark energy. It should  therefore come as no surprise that the DESI results have spawned a huge number of  papers, some geared towards phenomenological  models of dark energy, which will not be discussed  here,   and some devoted to the interpretation of the DESI results, which is
the focus of this paper. 

We will first briefly review the main DESI results  using numbers from DESI-DR2. The DESI team claim that the BAO measurements are in  tension with the base \LCDM\ cosmology preferred by \Planck\ at the $2.3\sigma$ level (implying odds against  DESI and  \Planck\ being compatible at about $50:1$). If the 
equation-of-state with redshift $z$ is parameterised as 
\begin{equation}
  w(z) = w_0 + w_a \left ( {z \over 1+z} \right ) ,  \label{equ:EoS1}
\end{equation}
\citep{Chevallier:2001, Linder:2003}, EvDE is preferred over \Planck\ \LCDM\ at the $3.1\sigma$ level if DESI BAO are combined with \Planck\ CMB measurements. Furthermore, if  constraints from the
magnitude redshift relation of Type Ia supernovae (SN) are included, the preference for EvDE lies between $2.8 - 4.2 \sigma$ depending on the choice of SN sample. The impression given by the DESI collaboration is that each time additional data, such as the CMB and SN are added to the analysis,  including the improvements in the  BAO measurements   from DR1 to DR2, the evidence for EvDE {\it strengthens}.
Needless to say, such a potentially important result has attracted a huge degree of interest and generated a diverse range of views \footnote{The literature on the DESI results has  become very large and is increasing day-by-day, so I will only reference key papers relevant to the discussion presented here.
For a survey of recent literature  see \cite{Giare:2024}.}.

\begin{figure*}
  \center
\includegraphics[width=160mm,angle = 0]{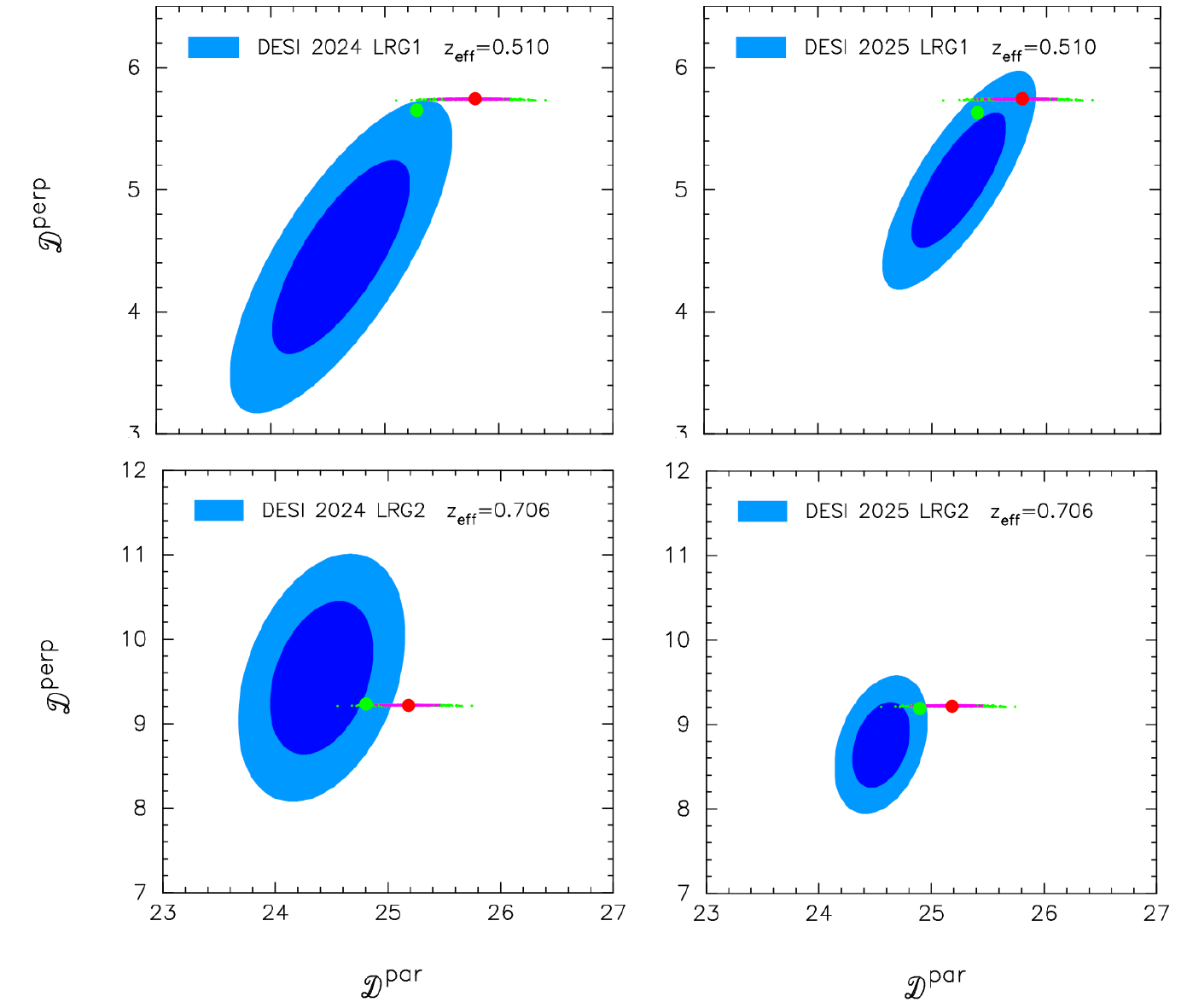}

  \caption{68\% and 95\% contours for the rotated BAO parameters \Dperp\ and \Dpar. We show results for the LRG1 and LRG2 surveys  for the 2024  DESI DR1 and 2025 DESI DR2 analyses. Samples from the  \Planck\ base \LCDM\ chains are shown by the dots: green for samples from the TT chains and purple for the TTTEEE chains. The spread along the \Dpar direction is determined to high accuracy by the value of the physical matter density parameter $\omega_m = \Omega_m h^2$. The red dots show the best fit values from the \Planck\ TTTEEE chains. 
  The green dots show the best fit values for the DESI+($\theta_*$, $\omega_b$, $\omega_{bc})_{\rm CMB}$ fits from DESI-DR1
and DESI-DR2.}

	\label{fig:DE_rot}

\end{figure*}

The purpose of this paper is to investigate whether the evidence for EvDE has actually strengthened between DESI-DR1 and DESI-DR2. We first note that the high significance levels \footnote{Note that although we quote the standard deviations from the DESI papers,   as discussed in Sect.~\ref{sec:adding_CMB},  we disagree with the statistical methodology and the impression that these these numbers are intended to convey.}  in favour of  EvDE come from
combining DESI BAO data with the DES5Y SN sample \citep{Sanchez:2024, Vincenzi:2024, DES5Y:2024}. In contrast, the Pantheon+ sample of \cite{Scolnic:2022} is  neutral to EvDE. The Union3 compilation of \cite{Rubin:2023} lies somewhere in between Pantheon+ and DES5Y. Thus different SN samples, which contain a high degree of
overlap with  each other, give different results. In a previous paper, I showed that inferences on EvDE from SN catalogues are very sensitive to the photometric accuracy of the SN  which must match to a precision of $\sim 0.01$ mag between low and high redshifts \citep{Efstathiou:2025}. This is
particularly acute for the DES5Y sample which is based on a large photometrically selected SN sample from DES (mainly at
redshifts $z \simgt 0.2$) but which is supplemented with a small number of nearby  SN from various
sources at redshifts $\simlt 0.1$\footnote{This has been noticed by a number of other authors: \cite{Gialamas:2025, Notari:2025, Huang:2025}.}. The DES team responded to this observation  using a simulation based investigation  to conclude that they 
saw  no reason to alter their analysis of the DES5Y SN compilation \citep{DESreply:2025}. This issue will not be discussed further in this paper, though it is important to note: (i) that there is no independent
check of the Beams with Bias Corrections methodology  \citep{Kessler:2017}  on which the simulations used in \citep{DESreply:2025} are based; (ii) Appendix A of
\cite{Efstathiou:2025} shows that the DES5Y SN
are in tension with the BAO from DESI DR1. As shown in the Appendix to this paper,  the tension with DES5Y SN gets worse with DESI DR2.

Given these differences, we ignore
the SN data in this paper\footnote{See  \citep{Cortes:2025} for a statistical argument to justify neglecting the SN.} and concentrate on the consistency between
DESI BAO and the CMB. This is an important question 
because, as noted above,  the DESI papers found that the combination of
DESI BAO and the CMB favours EvDE at the $3.1\sigma$
level: the addition of DES5Y SN, it is argued,   merely amplifies a trend in favour of EvDE that is already apparent from the combination of DESI BAO and the CMB.

To investigate this question, Sect.~\ref{sec:Rotations} presents a different way of viewing BAO measurements which 
allow a simple  and intuitive  test  of the \LCDM\ cosmology favoured by the CMB. This section addresses whether the BAO measurements from DESI-DR2 strengthen or weaken the evidence for \LCDM. Section~\ref{sec:adding_CMB} addresses the statistical methodology  employed by the DESI collaboration  with emphasis on  the combination of DESI BAO and the CMB. Our conclusions
are summarized in Sect.~\ref{sec:conclusions}. Appendix \ref{sec:appendix} compares the cosmological parameters inferred from DES5Y SN with those from DESI BAO and the CMB.

\section{A different angle on BAO}
\label{sec:Rotations}

BAO surveys measure the distance ratios $D_{\rm M}(z)/r_d$ and $D_H(z)/r_d$, where $r_d$ is the sound horizon, $D_{\rm M}$ is the comoving angular diameter distance and $D_H(z)$ is the Hubble distance given by\footnote{We assume a spatially flat Universe throughout this paper.}
\begin{subequations}
\begin{eqnarray}
D_{\rm M} (z) &=&  \int_0^z dz^\prime {c\over H(z^\prime)}, \label{equ:BAO1a} \\
D_H(z) & = &  {c \over H(z)}, \label{equ:BAO1b}
\end{eqnarray}
\end{subequations}
and $H(z)$ is the Hubble parameter. For a \LCDM\
universe with matter density parameter $\Omega_m$, $H(z)$ is given by
\begin{equation}
H(z) = H_0 \left [ \Omega_m (1+z)^3 + (1-\Omega_m)\right]^{1/2}. \label{equ:Hz}
\end{equation}

The errors on $D_{\rm M}(z)/r_d$ and $D_H(z)/r_d$
from BAO surveys such as DESI are set by the details
of the survey (e.g. redshift, volume, sampling rate, {\it etc.}) and by aspects of data processing (e.g. BAO reconstruction, estimator {\it etc.}). However,
it is evident from Eqs. \ref{equ:BAO1a} and \ref{equ:BAO1b} that $D_{\rm M}(z)/r_d$ and $D_H(z)/r_d$ are closely related and hence physics imposes strong constraints on these observables. This becomes clear if we analyse predictions of BAO distance ratios from  observations of the CMB.  
For example, Fig. 12 from \cite{Planck_Params_2018}
shows that in  \LCDM, the CMB requires $D_{\rm M}(z)/r_d$ and $D_H(z)/r_d$ to lie on  a narrow line
of negligible width compared to current BAO constraints. 

This behaviour is explained in Sect. 13.5 of \cite{Efstathiou:2021}.  In \LCDM, the
parameter combination\footnote{Where $h$ is the Hubble constant $H_0$ in units of $100\Hunit$.} $\Omega_m h^3$
is very tightly constrained by \Planck\ (since this is a
good proxy for the acoustic angular scale $\theta_*$). If
$\Omega_m h^3$ is fixed, the CMB BAO distance ratios are determined by the physical matter density $\omega_m = \Omega_m h^2$ (with a weak dependence on the baryon density
$\omega_b$ via the value of the sound horizon $r_d$,  which can be safely neglected in this paper).

It therefore follows that we can define two orthogonal distance ratios ${\cal D}^{\rm perp}$ and ${\cal D}^{\rm par}$:
\begin{subequations}
\begin{eqnarray}
{\cal D}^{\rm par} &=& (D_{H}/r_d) \sin \gamma(z) + (D_M/r_d) \cos \gamma(z), \label{equ:rot1a} \\
{\cal D}^{\rm perp} &=& (D_{H}/r_d) \cos \gamma(z) - (D_M/r_d) \sin \gamma(z), \label{equ:rot1b}
\end{eqnarray}
\end{subequations}
in which the \Planck\ degeneracy line is rotated by an angle 
$\gamma(z)$ so that it is horizontal  to high accuracy (i.e. parallel to the 
\Dpar\  axis). If the  measured value
of \Dperp\ lies a long way from the 
\Planck\ line, then this suggests a failure of the 
\LCDM\ cosmology which can be  quantified easily.

This is illustrated in Fig.~\ref{fig:DE_rot} which shows the DESI-DR1 and DESI-DR2 constraints for the
Luminous Red Galaxy  samples at  effective redshifts $z_{\rm eff}  = 0.51$ (LRG1) and $z_{\rm eff} = 0.706$
(LRG2). As discussed by the DESI collaboration, it is
only LRG1 and LRG2 that show some tension with \Planck\ \LCDM\ and so we focus on these two samples
here. $D_M/r_d$ and $D_H/r_d$ measurements of  DESI samples at higher redshift are  discussed later in this Section. For obvious reasons, we do not discuss the Bright Galaxy Sample at effective  redshift $0.295$, since the DESI team present only  an isotropised BAO distance ratio, $D_V/r_d$, for this
sample (which has a large error and agrees well  with the \Planck\ \LCDM\ cosmology).  One can see evidence for a  tension between
the DESI-DR1 posterior  of ${\cal D}^{perp}$ for the LRG1 sample and the \Planck\ expectation, but this becomes weaker in DESI-DR2.
Since the rotation angles $\gamma(z)$ are fixed by the effective redshift of the sample, the variable \Dperp\  provides  a powerful way of testing consistency between \Planck\ and BAO under the assumption of
\LCDM\ because the \Planck\ predictions are, to all intents and purposes, a delta function. There is no need to consider nuisance parameters, priors and other complexities. All one
has to do is to assess whether there is significant overlap
between the DESI posteriors for \Dperp\  and the \Planck\ values.

 \begin{figure*}
  \center
  \includegraphics[width=85mm, angle=-90]{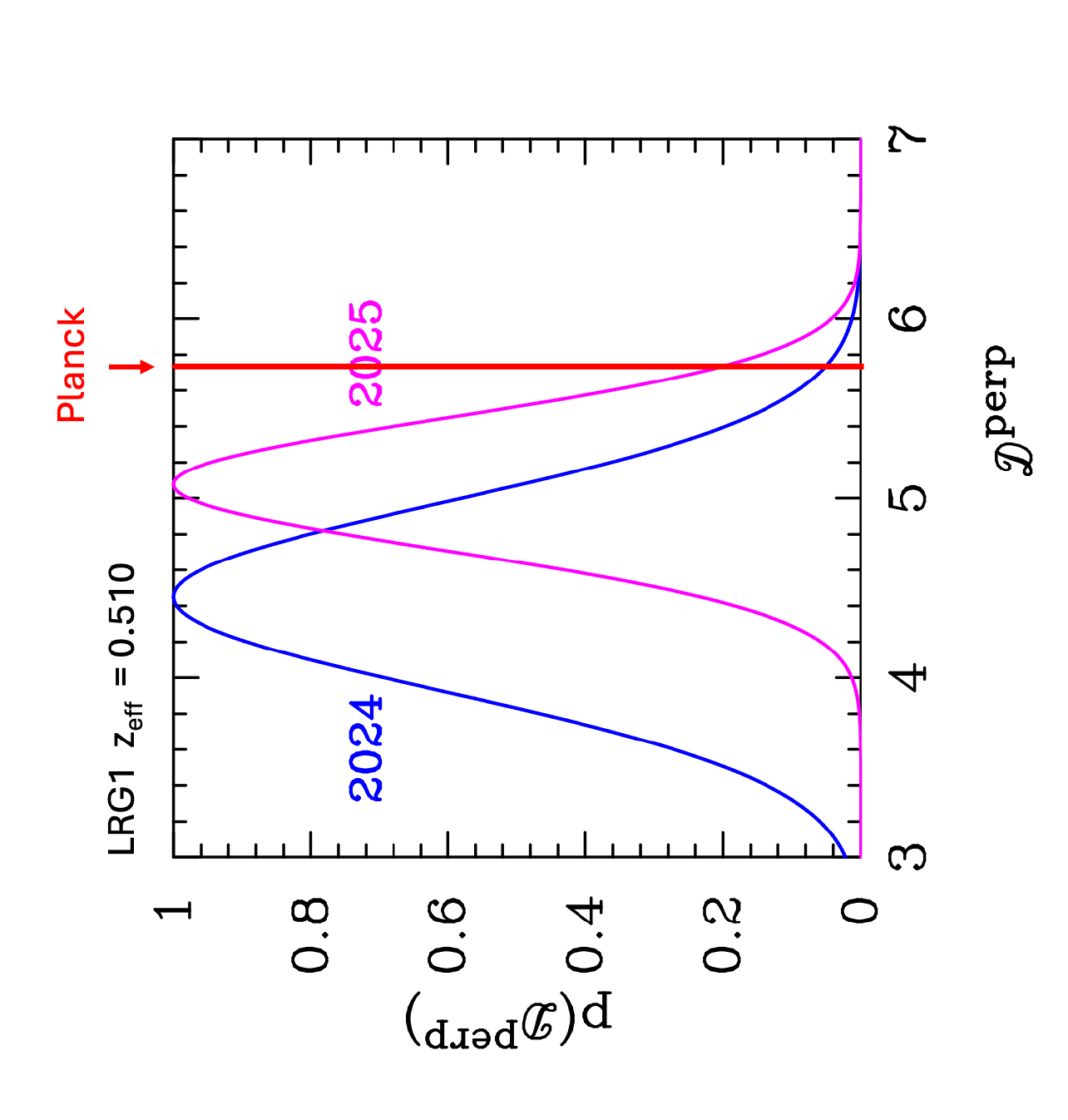} 
 \hspace{-0.3 truein} \includegraphics[width=85mm, angle=-90]{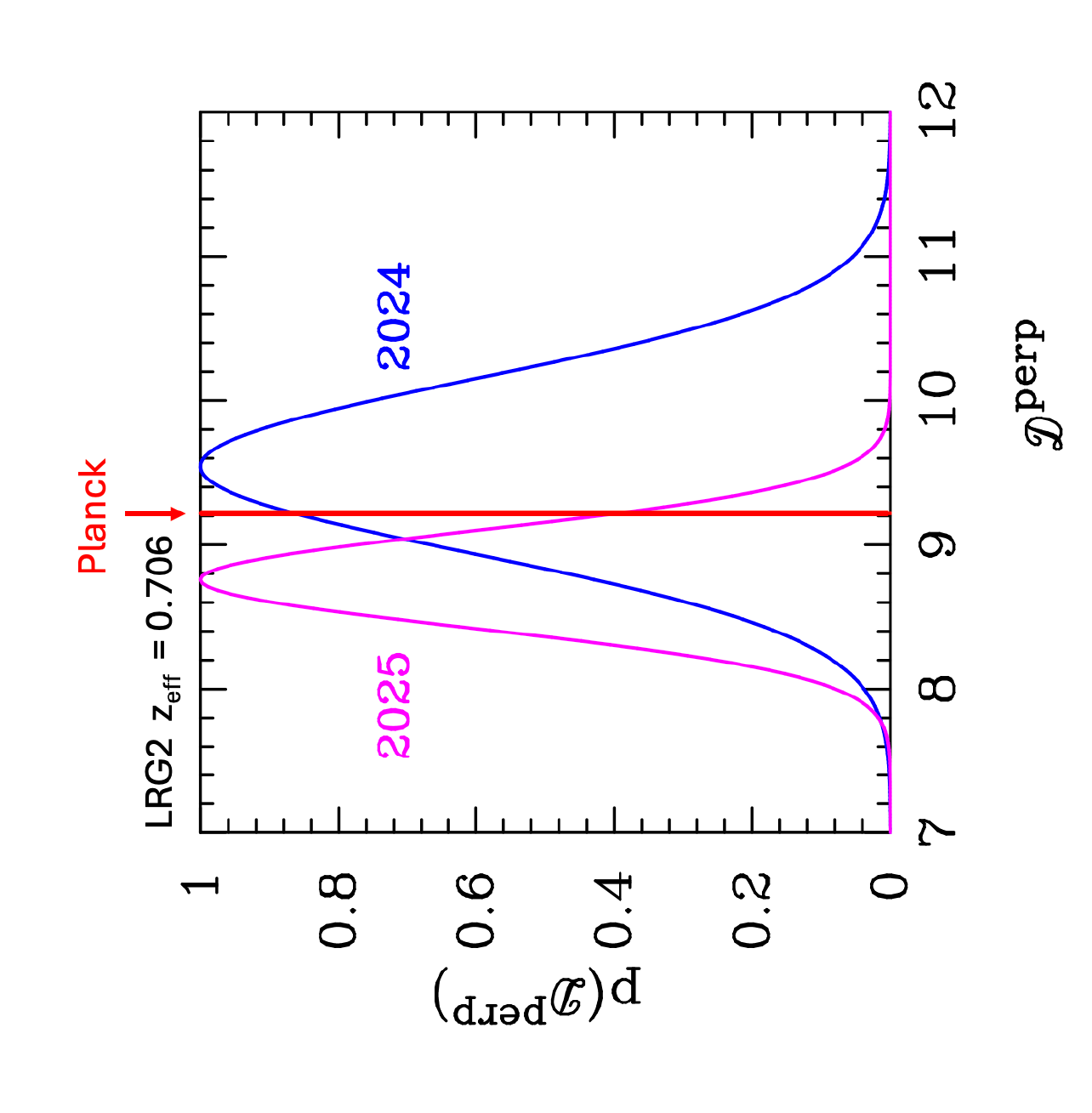}

  \caption{Posterior distributions of \Dperp\ for the DESI DR1 and DR2 LRG1 and LRG2 samples. The \Planck\ \LCDM\ values (which have negligible error)  are shown by the thick red lines. Note the $\sim 2.5\sigma$ tension between the \Planck\ value and  \Dperp\ measured from the DR1 LRG1 sample, which becomes less significant in DR2. }

	\label{fig:Dperp}

\end{figure*}

To perform this comparison, we marginalise the DESI constraints over ${\cal D}^{\rm par}$ to compute the posterior distrbution of \Dperp.
 Define $x = \Delta {\cal D}^{\rm perp}$,
$y = \Delta {\cal D}^{\rm par}$.  In the Gaussian approximation
\begin{equation}
 P(x, y) dx dy  \propto \exp \left ( -a_{11}x^2 - a_{22}y^2 - a_{12} xy \right )dx dy,  \nonumber
\end{equation}
marginalising over $y$ leads to  the distribution 
\begin{equation}
 P(x) dx   \propto \exp \left [ \left (-a_{11} + {a_{12}^2 \over 4 a_{22}} \right )  x^2 \right ]dx.  
\end{equation}
This gives the posterior distribution of $\Delta {\cal D}^{\rm perp}$.

\begin{figure}
  \center
 \hspace{0.05 in}  \includegraphics[width=80mm, angle=0] {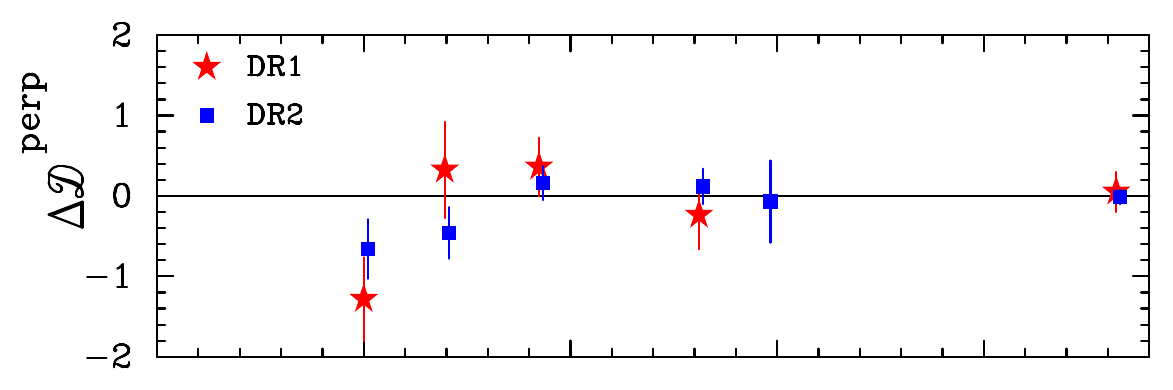} \\   \includegraphics[width=83mm, angle=0] {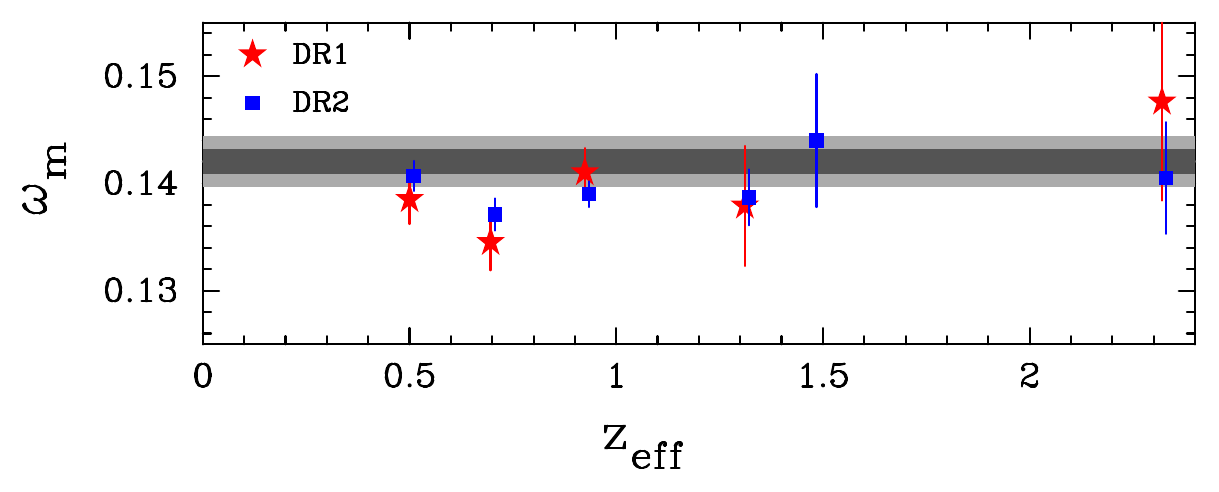}

  \caption{The upper panel shows the residual  $\Delta {\cal D}^{\rm perp}_{\rm DESI} - {\cal D}^{\rm perp}_{\rm  Planck}$ as a function of effective redshift for the DESI DR1 and DR2 samples. Note how the DR2 measurements shift closer to the \Planck\ values.  The lower panel shows the matter density parameter $\omega_m$ inferred from the DESI measurements of \Dpar. The horizontal bands show the $1$ and $2\sigma$ ranges allowed by \Planck\ for the base \LCDM\ cosmology. As in the upper panel, the DR2 results move closer to the expectations of the \Planck\ \LCDM\ cosmology.}

	\label{fig:Dperp_residual}

\end{figure}

The posteriors for ${\cal D}^{\rm perp}$ are shown in  Fig.~\ref{fig:Dperp} for the LRG1 and LRG2 samples.  It is clear from Fig.~\ref{fig:Dperp} that there is a high degree of overlap with the \Planck\ values at both redshifts and furthermore that the 2025 data {\it are more consistent} with \LCDM\ in comparison to the 2024 data. 
Table~\ref{tab:params} lists the rotations to create ${\cal D}^{\rm perp}$ and ${\cal D}^{\rm par}$ from the DESI measurements together with  various statistics including\footnote{Although I present p-values in Table~\ref{tab:params}, I will show in Sect.~\ref{sec:adding_CMB} that p-values can be highly misleading when they are applied to `quantify' more complex comparisons such as the evidence for EvDE.} two-tailed $p$-values for the comparisons between the DESI
posteriors for ${\cal D}^{\rm perp}$ and the \Planck\ values. The residuals of $\Delta {\cal D}^{\rm perp} = 
{\cal D}^{\rm perp} - {\cal D}^{\rm perp}_{\rm Planck}$ are plotted in the upper panel of Fig.~\ref{fig:Dperp_residual} as a function of redshift for all of the DESI samples. This illustrates the spectacular agreement between DESI DR2 and \Planck\ \LCDM\ over the entire redshift range covered by DESI. As in Fig.~\ref{fig:Dperp},  the two LRG samples come into better agreement with \Planck\ \LCDM\ with the improved BAO measurements of DR2. Evidently, there is no evidence
from the analysis of  ${\cal D}^{\rm perp}$ for any significant deviation from the \LCDM\ cosmology.

\begin{table*}

\begin{center}

  \caption{The first column lists the DESI sample and the second column lists the effective redshift. The third column gives the rotation angles $\gamma$ in radians used
  to define \Dpar\ and \Dperp. The fourth column gives the values of \Dperp\ computed from the \Planck\ TTTEEE \LCDM\ chains. These numbers have negligible error. The remaining columns list measurements of \Dperp\ from DESI DR1 and DESI DR2 (mean a posterior values) together with $\pm 1\sigma$ errors. The columns labelled $N_\sigma$ give the difference between the DESI and \Planck\  values of \Dperp\ in units of DESI standard deviation. The columns labeled $p$-values are computed from $N_\sigma$ assuming a two-tailed normal distribution.  }

\label{tab:params}

\smallskip

\begin{tabular}{l|c|c|c||c|c|c|c|c|c|} \hline 
    &  &  &  Planck    &  & DR1&  & & DR2&   \\ 
    &  &  &            & \multispan3$\quad\quad\quad\overbrace{\hspace{1.3 truein}}$ & \multispan3$\qquad\qquad\overbrace{\hspace{1.3 truein}}$  \\
    sample  &  $z_{\rm eff}$ & $\gamma$ & ${\cal D}^{\rm perp}$ & 
    ${\cal D}^{\rm perp}$ & $N_\sigma$ & p-value & ${\cal D}^{\rm perp}$ & $N_\sigma$ & p-value \\\hline
 LRG1 & $0.510$& \ \ $0.816$ & $5.74$ & $4.45\pm 0.52$ & \ \ $2.48$ & \  $0.013$  & \ $5.08 \pm 0.37$ & \ \ $1.78$ &\ $0.075$ \\
 LRG2  & $0.706$ &\ \ $0.500$ & $9.22$ & $9.54\pm0.60$ & $-0.53$ & $0.60$ & \ $8.76\pm0.34$ & \ \ $1.35$ & $0.16$\\
 LRG3+ELG1 & $0.934$ & \ \ $0.243$ & $11.76$ & $12.12 \pm 0.37$ & $-0.97$ & $0.33$ & $11.92 \pm 0.21$ & $-0.76$ & $0.45$ \\
 ELG2 & $1.321$ & \ \  $0.0274$ &$13.30$ & $13.06  \pm 0.43$ & \ \ $0.56$ & $0.58$ & $13.42 \pm 0.22$ & $-0.54$ & $0.59$  \\
 QSO & $1.484$ & $-0.0177$ & $13.42$ & - & - & -& $13.35 \pm 0.51$ & \ \ $0.14$ & $0.89$ \\
 Ly$\alpha$ & $2.33$ & $-0.0991$ &  $12.46$ &  $12.51 \pm 0.15$  & $-0.33$ & $0.74$ & $12.448 \pm 0.091$ &\ \  $0.14$ & \ \  $0.89$ \ \   \\\hline
\end{tabular}
\end{center}
\end{table*}

\begin{table*}

\begin{center}

  \caption{The first two columns are as in Table~\ref{tab:params}. The third column gives the slope $\beta$ of Eq.~\ref{equ:omega1} at each redshift necessary to convert measurements of \Dpar\ to constraints on $\omega_m$ according to the \Planck\ \LCDM\ cosmology. The columns labelled \Dpar\  list the  mean a posterior values of \Dpar\ from DESI DR1 and DR2 together with $\pm 1\sigma$ errors. The columns labelled $\omega_m$ convert the measurements of \Dpar into measurements of $\omega_m$ using Eq.~\ref{equ:omega1}.}

\label{tab:parallel}

\smallskip

\begin{tabular}{l|c|c|c|c|c|c|c|} \hline 
    &  &  &  Planck &      \multispan2{\hspace{-0.08 truein} DR1} & \multispan2{\hspace{-0.10 truein} DR2}   \\ 
    &  &  &            & \multispan2$\quad\quad\quad\overbrace{\hspace{0.9 truein}}$ & \multispan2$\qquad\quad\overbrace{\hspace{0.9 truein}}$  \\
    sample  &  $z_{\rm eff}$ & $\beta$ & ${\cal D}^{\rm par}$ & ${\cal D}^{\rm par}$ &
    $\omega_m$  &   ${\cal D}^{\rm par}$ & $\omega_m$ \\\hline
 LRG1 & $0.510$& $0.00836$ & $25.776 \pm 0.126$  & $25.35 \pm 0.26$ & $0.1385\pm 0.0022$ & $25.62 \pm 0.17$ &   $0.1407 \pm 0.0014$  \\
 LRG2  & $0.706$ & $0.00921$ &$25.170\pm0.115$ & $24.36 \pm 0.29$ & $0.1345 \pm 0.0026$ & $24.63 \pm 0.16$ & $0.1371 \pm 0.0015$ \\
 LRG3+ELG1 & $0.934$ & $0.00934$ &$25.549\pm 0.113$ & $25.44 \pm 0.25$ & $0.1410 \pm 0.0023$ & $25.22 \pm 0.13$ & $0.1390 \pm 0.0012$ \\
 ELG2 & $1.321$ & $0.00919$& $28.425 \pm 0.115$ &$27.98\pm 0.61$ & $0.1379 \pm 0.0056$  & $28.08 \pm 0.28$ & $0.1387 \pm 0.0026$  \\
 QSO & $1.484$ & $0.00919$ & $30.012\pm 0.115$ & - & - &   $30.24 \pm 0.67$ & $0.1441 \pm 0.0062$ \\
 Ly$\alpha$ & $2.33$ & $0.00975$ & $38.114\pm 0.108$ &  $38.69 \pm 0.94$ &  $0.1476 \pm 0.0092$  & $37.95 \pm 0.53$ & $0.1405 \pm 0.0052$  \\\hline
\end{tabular}
\end{center}
\end{table*}

What about the parallel direction? To compute the distribution of ${\cal D}^{\rm par}$ conditional on the \LCDM\ cosmology, we need to slice the DESI likelihood with ${\cal D}^{\rm perp}$ fixed at the \Planck\ value (since to a good approximation the  slice can be assumed to be infinitely thin). This shifts the mean of the distribution of ${\cal D}^{\rm par}$:
\begin{eqnarray}
 & & P({\cal D}^{\rm par})d{\cal D}^{\rm par} \propto   
  \exp  {\bigg [}  - a_{22}  {\bigg  (} {\cal D}^{\rm par} - {\cal D}^{\rm par}_{\rm DESI}   + \hspace{0.3 truein} \nonumber  \\  
& & \hspace{0.8 truein} \left.   \left. {a_{12} \over 2 a_{22}} ({\cal D}^{\rm perp}_{\rm Planck} - {\cal D}^{\rm perp}_{\rm DESI}) \right ) \right ]   d{\cal D}^{\rm par} .  \label{equ:Dpar}
\end{eqnarray}
Significantly, this shift in the mean generally brings the BAO distributions into closer agreement with the \Planck\
\LCDM\ cosmology. This can be seen in the orientations of the ellipses in Fig.~\ref{fig:DE_rot}, which are tilted towards the \Planck\ best fit cosmology. (This tendency is also apparent in Fig.~14 of \cite{Adame:2025}).

Results for the parallel statistics are summarized in Table~
\ref{tab:parallel}. Using the \Planck\ \LCDM\ TTTEEE chains, the deviation from the \Planck\ best-fit value of \Dpar\
(listed in the fourth column) is well approximated by the linear relation\footnote{Note that the DESI papers make a distinction between $\omega_{bc}$, the physical density in baryons and cold dark matter, and the quantity $\omega_m= \omega_{bc} + \omega_m$ used in this paper. The two density parameters differ by the contribution from massive neutrinos. For a minimal-mass normal
hierarchy the sum of neutrino masses can be accurately approximated as  a single  massive eigenstate of mass
$m_\nu = 0.059\ {\rm eV}$ \citep{neutrino_mass}  giving  $\Omega_\nu h^2 = 0.00063$. It makes a small difference whether one correlates the \Planck\ \LCDM\ predictions with $\omega_m$ or $\omega_{bc}$. }
\begin{equation}
\omega_m = 0.14205 + \beta({\cal D}^{\rm par} - {\cal D}^{\rm par}_{\rm Planck}). \label{equ:omega1}
\end{equation}
Thus, the DESI measurements of \Dpar\ (listed  in the fifth and seventh columns) can be converted into constraints on
the matter density parameter $\omega_m$ (listed in the sixth and eighth columns).

The DESI constraints on $\omega_m$ are plotted in the lower panel of Fig.~\ref{fig:Dperp_residual}, together with the \Planck\ 1 and 2 $\sigma$ constraints on $\omega_m$ ($\omega_m = 0.14205 \pm 0.00123$). As in the discussion of \Dperp, the DESI changes from DR1 to DR2 move the points  {\it closer} to the \Planck\ value. There is mild tension in the value of $\omega_m$ inferred
from the LRG2 sample: for DR2, the value of $\omega_m$ is about $2.6\sigma$ lower than expected from \Planck\ (adding the errors from both measurements in quadrature). Averaging over all redshifts, the best fit values of $\omega_m$ are $0.1384 \pm 0.0015$ from DR1  and
$0.13909 \pm 0.00073$ from DR2 differing from the \Planck\ value by $\sim 1.9$ and $2.1 \sigma$ respectively. For reference, the recent results from the Atacama Cosmology Telescope (ACT)  \citep{ACT_DR6:2025} 
for the P-ACT combination (TTTEEE \Planck\ power spectra at $\ell <1000$ and ACT spectra at higher multipoles)  give $\omega_m = 0.1425 \pm 0.0012$, i.e. very close to the \Planck\ result.

In summary, the analysis of this Section presents a different way of viewing BAO  that makes it easy to assess whether the measurements are compatible
with the \LCDM\ cosmology. In addition, we have shown that
DESI-DR2 measurements are substantially more consistent with the  \Planck\ \LCDM\ cosmology
in the perpendiclular  direction and only marginally worse in the parallel direction compared to DESI-DR1,  despite the  large increase in statistical power of DESI-DR2 over
DESI-DR1.   It is difficult to argue against the results shown in Figs~\ref{fig:Dperp} and \ref{fig:Dperp_residual} that DESI DR2 BAO come into better consistency with \Planck\ \LCDM.

\begin{figure*}
  \center
   \includegraphics[width=80mm, angle=0] {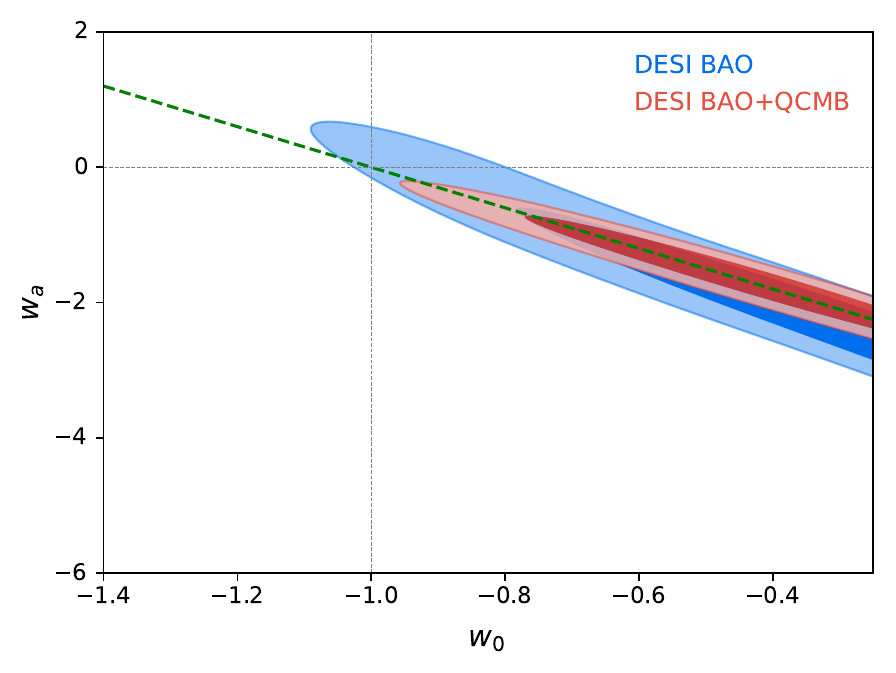}    \includegraphics[width=80mm, angle=0] {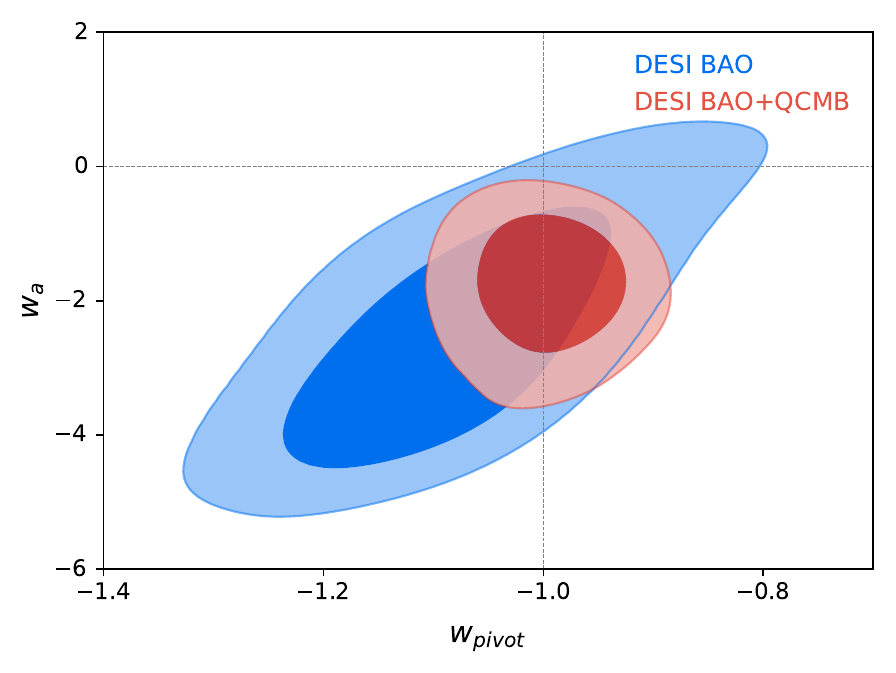}

  \caption{The left hand panel shows 68\% and 95\% contours for the marginalized posterior parameters in the $w_0-w_a$, $w_0-\Omega_m$ and
    $w_a-\Omega_m$ planes for the Pantheon+  (red) and DES5Y (blue) SN compilations. The dashed lines show the values for
  the best fit \Planck\ \LCDM\ cosmology. The green dashed line shows the relation of Eq.~\ref{equ:piv2} determined by fixing $w_{\rm piv} = w(z_{\rm piv}) = -1$ at a pivot redshift $z_{\rm piv} = 0.5$. The right hand panel shows the constraints on $w_{\rm piv}$ and $w_a$, illustrating the tight constraints on $w_{\rm piv}$ and the decorrelation of $w_{\rm piv}$ from $w_a$.}

	\label{fig:w0wa}

\end{figure*}

Although \LCDM\ appears to be a good description of both the CMB and the BAO
measurements, the DESI collaboration argue that the situation changes with 
the  introduction of  EvDE. Including the additional parameters $w_0$ and $w_a$ from Eq.~\ref{equ:EoS1}, they claim to find  a solution that `is preferred over \LCDM\ at $3.1\sigma$ for the combination of DESI BAO and CMB data'. The veracity of this statement is the subject of the next Section.

\section{Combining BAO and the CMB: another rotation}
\label{sec:adding_CMB}

In this Section, I will  follow the DESI collaboration and use
a compressed CMB  likelihood consisting of three parameters ($\theta_*$, $\omega_b$, $\omega_{m}$, see Eqs A1 and A2 of DESI-DR2).
I will denote this likelihood QCMB. The simplified  CMB likelihood is perfectly adequate for our purposes  and is insensitive to physical processess that have nothing to do with EvDE that can lead, via the late-time integrated Sachs-Wolfe (ISW) effect \citep{Sachs:1967, Rees:1968}, to small shifts that can be mistaken for EvDE (though always at low statistical significance). Analysis of such effects will be deferred to another paper.

The left hand panel of Fig.~\ref{fig:w0wa} shows the constraints from DESI-DR2 if $w_0$ and $w_a$ are allowed
to vary as additional  parameters to those of  \LCDM\footnote{We use the {\tt MULTINEST} nested sampler \citep{Feroz:2009, Feroz:2011} to produce parameter chains in this paper. The choices of priors are as described in \cite{Efstathiou:2025}.}. Following DESI-DR2 we denote these models as $w0wa$CDM. Adding the QCMB likelihood leads to
the red contours in Fig.~\ref{fig:w0wa}. These contours agree well with those in Fig.~14 of DESI-DR2 (which are actually close to those using the full \camspec\ TTTEEE \Planck\ likelihood). Now let us follow the interpretation given in DESI-DR2. The DESI team calculate $\Delta \chi^2_{\rm MAP}$, which is
the difference in $\chi^2$ at the  maximum a posteriori likelihood of the w0waCDM model and the maximum
a posterior likelihood  for \LCDM. In this case,  $\Delta \chi^2_{\rm MAP} = -8.0$. Thus adding two additional parameters leads to improved fits to the data. Assuming a 
$\chi^2$ distribution with two degrees of freedom, the  $\Delta \chi^2_{\rm MAP}$ gives a p-value of $0.0183$. This p-value is then converted into
`standard deviations' by matching to the two-tailed p-value
of a normal distribution. In this case, the result is summarized as a $2.4 \sigma$ preference for EvDE. The impression given by these numbers is that there
is a preference for  w0waCDM and the  odds  are stacked against  \LCDM\ by  about 55:1.
Is this a reasonable conclusion?

One can see from Fig.~\ref{fig:w0wa} that the red contours are quite narrow and oriented towards the \LCDM\ point
$w_0=-1$, $w_a = 0$. (Note that the main effect of the  more accurate BAO measurements from DESI-DR2 is to make
the  contours narrower perpendicular to the main degeneracy direction). It is clear from this Figure that $w_0$ and $w_a$ are highly correlated. The green dashed line in the left hand panel of Fig.~\ref{fig:w0wa} is computed as follows. Define $w_{\rm piv}$ to be the value of the equation-of-state parameter at a pivot redshift, $z_{\rm piv}$:
\begin{equation}
w_{\rm piv} = w_0 + w_a \left ( z_{\rm piv} \over 1 + z_{\rm piv}\right ).    \label{equ:piv1}
\end{equation}
Now fix $w_{\rm piv}= -1$ at a redshift at which $w(z)$ is 
determined  accurately by the data, say $z_{\rm piv} = 0.5$\footnote{A more objective way of chosing $z_{\rm piv}$ is to match the degeneracy direction of the combined
DESI BAO+CMB likelihood, decorrelatong $w_{\rm piv}$ from $w_a$. It is clear from Fig.~\ref{fig:w0wa} that this is achieved for
$z_{\rm piv} \approx 0.5$.}.
In this case, 
\begin{equation}
 w_0 = -1 - w_a \left ( z_{\rm piv} \over 1 + z_{\rm piv}\right ),    \label{equ:piv2}
\end{equation}
which is plotted as the green dashed line in Fig.~\ref{fig:w0wa}\footnote{The dashed line defined by Eq.~\ref{equ:piv2} is quite close to the CMB geometric degeneracy direction in which models with $w_0$, $w_a$ have the same angular diameter distance to the last scattering surface. It is this unfortunate alignment of the degeneracy directions that make it difficult to constrain EvDE using BAO and the CMB.}.  Evidently, if we choose to reparameterize $(w_0, w_a)$ in terms of $w_{\rm piv} \equiv
w(z_{\rm piv})$, the parameters $w_{\rm piv}$ and $w_a$ will decorrelate \citep{Cortes:2024}.  This is illustrated in the right hand panel of Fig.~\ref{fig:w0wa}. With this rotation, the maginalised EvDE  parameters  are:
\begin{subequations}
\begin{eqnarray}
& &\hspace{-0.28 truein}w_{\rm piv} = -1.08 \pm 0.10, \ w_a = -2.81 \pm 1.38, \ {\rm DESI}, \\
& &\hspace{-0.28 truein}w_{\rm piv} = -0.996 \pm 0.046, \  w_a = -1.78 \pm 0.79, \ {\rm DESI+QCMB}. \nonumber \\
& & 
\end{eqnarray}
\end{subequations}
For both data combinations, $w_{\rm piv}$ is consistent with the \LCDM\ value of $w=-1$ and it is  accurately
determined by the data. The $\Delta \chi^2_{\rm MAP}$ is
of course unchanged and so according to the interpretation of the  DESI collaboration the odds in favour of EvDE remain at about 55:1 and the fact that  $w(z)$ agrees with the \LCDM\ value at a redshift where it is well determined by the data is a coincidence. The peculiarity of this conclusion has been pointed out by \cite{Cortes:2025}.

The interpretation becomes even more peculiar if we fix
$w_{\rm piv}$ = -1 and test a model of EvDE which has only one parameter $w_a$. In this case, $\Delta \chi^2_{\rm MAP}$
does not change because the maximum a posterior value of $w_a$ lies accurately along the degeneracy direction of
Eq.~\ref{equ:piv2}. However, this model has only one 
parameter and so the odds in its favour increase to $212:1$
(increasing the preference for EvDE from $2.4\sigma$ to $2.8 \sigma$).  So is this model  more likely  than the
$w_0, w_a$ parameterization? Intuitively, the one parameter model seems more contrived {\it but crucially there is no measure of contrivedness  in  the DESI statistical methodology.}

The Bayesian solution to this type of problem is well known. One computes the Evidence (the probability of the data $D$ given the model $M$):
\begin{equation}
 E(M) = \int d\pmb{$\theta$}P(D \vert\pmb{$\theta$} M)  \pi ( \pmb{$\theta$} \vert M), \label{equ:bayes1} 
\end{equation}
where  $\pi ( \pmb{$\theta$} \vert M)$ is the prior distribution of model parameters \pmb{$\theta$}\  and 
$P(D \vert\pmb{$\theta$} M)$ is the likelihood of the parameters under model $M$. If we have two models, $M_1$ and $M_2$, the Bayes factor $B_{12} = E(M_1)/E(M_2)$ provides a measure with which to discriminate between the
models. Critically, the Evidence requires specification of the prior distribution $\pi ( \pmb{$\theta$} \vert M)$.
I have previously written about the limitations of Bayesian Evidence applied to cosmology \citep{Efstathiou:2008}. This is not because Evidence is the wrong thing to calculate, but rather that  in cosmology we often have no guidance on how to choose priors.  Hence there is no objective way of computing Bayes
factors that everybody can agree on.  The ideal situation
in Bayesian statistics is for the data to overwhelm the priors. This happens in the \Planck\ analysis of \LCDM\ for 
parameters such as $\omega_b$, $\omega_c$ etc. But the situation is very different if the \LCDM\ model is extended, for example,  to include curvature $\Omega_K$. In this case, the CMB displays a strong geometric degeneracy and so the interpretation  of CMB data alone becomes  very sensitive to priors. This 
is an especially difficult problem because unlike  parameters such as $\omega_b$ and $\omega_c$, there are good theoretical reasons for why there may be an attractor  driving $\Omega_K$ to a particular value. Thus, in inflationary cosmology $\Omega_K = 0$ is a special case. 
Fortunately the geometrical degeneracy is broken decisively by other data, 
especially BAO, leading to the strong constraint $ \Omega_K =
0.0004\pm 0.0018$, consistent with inflationary cosmology
\citep[for further  discussion of these issues see][]{Efstathiou:2020}.

As with $\Omega_K$, a strong geometrical degeneracy re-emerges if the \LCDM\ cosmology is extended to include the dark energy parameters  $w_0$ and  $w_a$. The CMB constraints on these parameters then stratify along lines of constant angular diameter distance
to recombination with only a weak dependence coming from the ISW effect to break the degeneracy \citep[see e.g. Fig.~30 of][]{Planck_Params_2018}. However, unlike the case of $\Omega_K$,  BAO constraints follow almost the same degeneracy direction as the CMB  (see Fig.~\ref{fig:w0wa}) and so do not decisively break the geometrical degeneracy. This means that the combination of CMB+BAO is mired in the swampland
of $\sim 2-3 \sigma$ statistics,  where the interpretation is strongly dependent on
choices of priors (the bane of cosmological inference). 

How do we choose priors in the case of EvDE? This is even more difficult than choosing a prior for $\Omega_K$. We have no theoretical understanding of the cosmological constant problem \citep{Weinberg:1989} apart from possible anthropic
explanations\footnote{Which this author believes are ill-posed in the context of inflationary cosmology.} \citep{Weinberg:1987,Efstathiou:1995}. In fact, prior to
the discovery of dark energy \citep{Riess:1998, Perlmutter:1999}, the prior 
$\Lambda =0$ favoured by the majority of theorists would have been wide of the mark. The case of EvDE poses even more difficulties than $\Lambda$. To display interesting dynamical behaviour at late times, a scalar field model of EvDE must involve a very small mass of $\sim 10^{-32} {\rm eV}$. This requires some sort of shift symmetry to protect the small mass from radiative corrections. In addition, to produce 
$w_{\rm piv}$ close to $-1$, as demanded by the data (Eq.~\ref{equ:piv2}), requires fine tuning of the potential. Thus although models can be constructed that map onto the region of $w_0$, $w_a$ space allowed by the data, 
\citep{Shlivko:2024, Wolf:2025a, Wolf:2025b, Shlivko:2025}, the parameters involve unexplained fine-tunings. It therefore seems reasonable to this author that models of EvDE should be given a {\it substantial} penalty compared to \LCDM. How big a  penalty is entirely subjective\footnote{As is clear from the arguments presented by \cite{Brandenberger:2025}.}, but requiring odds of order $10^6:1$ or more (i.e. about  the  threshold  of $5\sigma$ commonly applied in  particle physics) would seem  reasonable before claiming evidence for new physics. 

\section{Conclusions}
\label{sec:conclusions}

The first part of this paper should be uncontroversial. It presents a different way of looking at the DESI BAO data by constructing orthogonal linear combinations of $D_{H}/r_d$ and $D_{M}/r_d$, \Dperp and \Dpar, 
designed to simplify tests of the \Planck\ \LCDM\ cosmology. The analysis is not
merely a trivial rotation, however, because in comparing with \LCDM\   we can impose constraints on the DESI measurements to account for regions of parameter space that are inaccessible to a \LCDM\ cosmology. 

The results for \Dperp\ plotted in the upper panel of Fig.~\ref{fig:Dperp} for the LRG1 and LRG2 samples (which in DR1 showed  the strongest discrepancy with \Planck\ \LCDM)  demonstrate unambiguously that DESI DR2 measurements move closer towards the expectations of \Planck\ \LCDM. In addition, the changes in DR2 bring the BAO measurements for LRG1 and LRG2 into closer agreement with the earlier SDSS/BOSS measurements summarized by \cite{Alam:2017}.
The improvement in the consistency in \Dperp\ between DESI BAO and \Planck\ \LCDM\ extends over the entire redshift range of the DESI measurements.  We find similar consistency for the orthogonal statistic \Dpar\ as shown in the lower panel of Fig.~\ref{fig:Dperp}. Despite the increase in statistical power of DESI  DR2, the measurements move closer to \Planck\ \LCDM\ {\it with no evidence to support EvDE}. This is a triumph for experiment and theory, for which the DESI collaboration deserves enormous credit.

I suspect that the second part of the paper will be considered more controversial, though it is based on well-worn arguments in Bayesian statistics \citep[see e.g.][]{jaynes03, MacKay2003}. The DESI collaboration argue that if DESI BAO are  combined with \Planck\ CMB,  EvDE is preferred over \LCDM\ at the $3.1\sigma$ level with the clear implication that the odds in favour of EvDE are about 500:1. But probability represents a degree of belief and is necessarily subjective. In physics, and especially in cosmology, there are rarely symmetry principles or other considerations to objectively fix priors. Hence there is no objective measure of
the odds in favour of EvDE.  

To  believe that the data point to  EvDE, one needs to:
\begin{itemize}
\item[(i)] believe that  any model in the $w_0$, $w_a$ plane\footnote{Apart from imposing a prior $w_0 + w_a < 0$  to ensure matter domination at high redshift.} has the same prior probability as $\Lambda$;

\item[(ii)] believe that  out of all possible models for EvDE, those favoured by the data have  $w_{\rm piv}$ at $z_{\rm piv}=0.5$ very close to $-1$ 
(Eq.~\ref{equ:piv1}) and that it is a  coincidence  that the density of dark energy  happens to peak at a redshift at which the equation-of-state is accurately determined by the data;

\item[(iii)] believe that the heuristic $\Delta \chi^2_{\rm MAP}$ statistic\footnote{DESI-DR1 and DESI-DR2 also quote changes in the Deviance Information Criterion $\Delta{\rm DIC}$, but this statistic is based on the sampling chains with uniform priors in $w_0$, $w_a$. $\Delta {\rm DIC}$  adds nothing new and is in fact well approximated by $\Delta \chi^2_{\rm MAP} = \Delta{\rm DIC} - 4$. \cite{Lodha:2025} quote Bayes factors for some models, but again assuming the same priors as those used in the sampling chains. } accurately accounts for overfitting of the data and gives a measure of the odds against \LCDM;

\item[(iv)] disregard the fact that with the very substantial increase in statistical power of DESI BAO from DR1 to DR2 there is almost no change in the `significance' levels  quoted by the DESI collaborations in favour of EvDE. (Noting that  in this paper we find that DR2 gives substantially better  consistency  with \LCDM\ than DR1 for the  \Dperp\ statistic); 

\item[(v)] discount the fact that the inclusion of powerful new CMB data  from ACT \citep{ACT_DR6:2025} in place of, or in addition to, \Planck\ does not significantly strengthen the evidence for EvDE and that some ACT+\Planck\ data combinations actually weaken the evidence for EvDE \citep{Garcia-Quintero:2025}.
\end{itemize}

A quantity such as $\Delta\chi^2_{\rm MAP}$ cannot, and should not,  be used to imply a
probability. Degrees of belief are subjective quantities  and so there will be differences of opinion unless data so overwhelm reasonable choices of priors that a consensus emerges.

What about the SN?  As noted in the introduction, The impact of adding SN to DESI and BAO is strongly dependent on which SN compilation is used.  Until the reasons for these differences are understood, it is  dangerous to place much faith in the SN \citep[see also][]{Cortes:2025} and highly misleading  to present results for one SN compilation while ignoring the others  \citep[see e.g.][]{Shajib:2025}. I have previously argued that there may be a systematic 
mismatch in the photometry of nearby ($z \simlt 0.1$) SN and those at higher redshift that  drives cosmological parameters inferred from the  DES5Y catalogue into regions of parameter space that are strongly disfavoured by other data \citep{Efstathiou:2025}.  Appendix A provides an update and shows that the tension between parameter constraints from DES5Y SN and those determined from DESI BAO and the CMB gets worse from DR1 to DR2. This should raise serious concerns about whether it is reasonable to combine DES5Y SN with BAO and the CMB.

\section{Acknowledgements}
I thank the Leverhulme Foundation for the award of a Leverhulme Emeritus Fellowship. I am grateful to Arnaud de Mattia who  has checked the calculations given in this paper using $\omega_{bc}$ in place of $\omega_m$, finding small but insignificant
differences.  I thank Marina Cort\^es and Andrew Liddle for comments on the manuscript.

\section*{Data Availability} 

No new data were generated or analysed in support of this research.

\appendix
\section{Comparison of cosmological parameter constrains from DES5Y SN compared to DESI DR2 BAO and QCMB}
\label{sec:appendix}
\begin{figure*}
  \center
   \includegraphics[width=60mm, angle=-90] {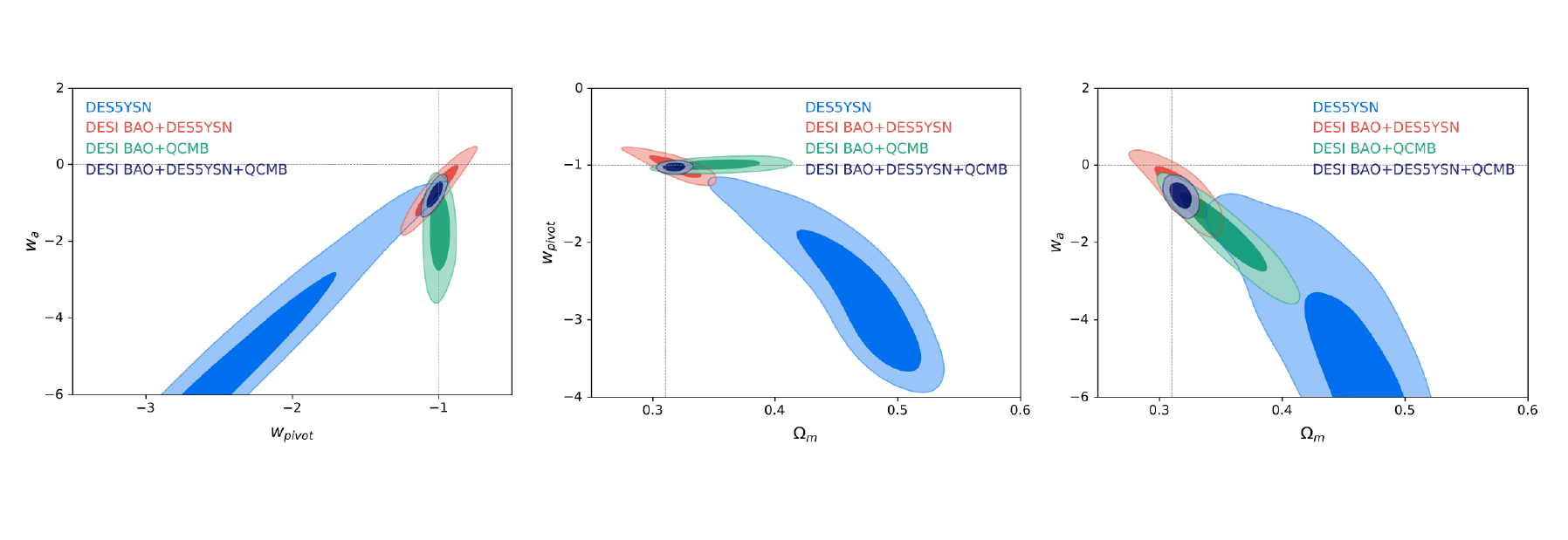}    

  \caption{68\% and 95\% contours for the marginalized posterior parameters in the $w_{\rm piv}-w_a$, $w_{\rm piv}-\Omega_m$ and
    $w_a-\Omega_m$ planes for DES5Y (blue) SN compilations. The remaining contours
    show DES5Y combined with QCMB (green), DESI BAO combined with DES5YSN (red), and DESI BAO combined with DES6YSN and QCMB (dark blue). The dashed lines show the values for the best fit \Planck\ \LCDM\ cosmology.}

	\label{fig:DES5Y}

\end{figure*}

Figure~\ref{fig:DES5Y} compares cosmological constraints from the DES5Y SN catalogue in the space of
$w_{\rm piv}$. $w_a$ and $\Omega_m$. As discussed in \cite{Efstathiou:2025}  the DES5Y contours are skewed to high values of $\Omega_m$ that are disfavoured by  other  data. In the central panel there is a clear separation between $w_{\rm piv}$ determined from  the DES5Y SN and the constraints from DESI BAO, which favour $w_{\rm piv}$ close to $-1$ to high accuracy. One can clearly see how the
DES5Y SN data drag the rest of the data away from the cosmological constant point. For DESI BAO + DES5Y SN + QCMB, we find $\Delta \chi^2_{\rm MAP} = 17.1$
relative to \LCDM. Following the statistical methodology of the DESI collaboration, this gives a $p$-value of $0.0001297$ corresponding to a $3.8\sigma$ `preference' for EvDE over \LCDM. 

\bibliographystyle{mnras}
\bibliography{DEpaper} 

\end{document}